# An Overview of Cyber Threats, Attacks, and Countermeasures on the Primary Domains of Smart Cities

**Vasiliki Demertzi[1], Stavros Demertzis[2] and Konstantinos Demertzis[3*]**

[1] Computer Science Department, School of Science International Hellenic University Kavala Campus, 65404 Kavala, Greece; vademer@teiemt.gr
[2] School of Spatial Planning and Development Faculty of Engineering, Aristotle University of Thessaloniki 54124 Thessaloniki, Greece; demertzs@plandevel.auth.gr
[3] School of Science & Technology, Informatics Studies, Hellenic Open University, 26335 Patra, Greece; demertzis.konstantinos@ac.eap.gr
*Correspondence: demertzis.konstantinos@ac.eap.gr

**Abstract:** A smart city is a place where existing facilities and services are enhanced by digital technology to benefit people and companies. The most critical infrastructures in this city are interconnected. Increased data exchange across municipal domains aims to manage the essential assets, leading to more automation in city governance and optimization of the dynamic offered services. However, no clear guideline or standard exists for modeling these data flows. As a result, operators, municipalities, policymakers, manufacturers, solution providers, and vendors are forced to accept systems with limited scalability and varying needs. Nonetheless, it is critical to raise awareness about smart city cybersecurity and implement suitable measures to safeguard citizens' privacy and security because the cyber threats seem to be well-organized, diverse, and sophisticated. This study aims to present an overview of cyber threats, attacks, and countermeasures on the primary domains of smart cities (smart government, smart mobility, smart environment, smart living, smart healthcare, smart economy, and smart people) to present information extracted from state-of-the-art to policymakers to perceive the critical situation and, at the same time, to be a valuable resource for the scientific community.

**Keywords:** Smart City; Cyber Threats; Cyber Attacks; Smart Government; Smart Mobility; Smart Environment; Smart Living; Smart Healthcare; Smart Economy; Smart People;

## 1. Introduction

The smart city [1] is an ecosystem that offers various e-government services, ensuring the seamless access of participating citizens to these services. At the same time, through an integrated analysis program of the information collected, you promote the optimal use of available resources, the improvement of urban space, and its other administrative services. A city can be considered smart when traditional infrastructure and investment in human resources support sustainable economic development and high quality of life based on ICT. In this spirit, a smart city can connect its built environment, which is also its natural capital, with society, businesses, and human resources to develop better services and infrastructure for its perpetual sustainability [2].

As said, the role of ICT is to provide intelligent management tools that will unite and strengthen the networks of people, infrastructure, companies, and generally available resources with the aim of sustainable economic development, high quality of life, and general well-being for the vast majority of citizens. Therefore, a smart city is a city that tries to face and solve public issues with the help of technology but based on a participatory process between multiple stakeholders, with prudent management of natural reseizures above all, through participatory action and active participation of citizens, preventing and many times eliminating social exclusion. Thus, smart cities ensure a networked urban society, which enjoys the benefits of intelligent management of its affairs with minimal financial, administrative, and social costs [3].

Here it is also important to mention that a city cannot be considered smart if useful, up-to-date, and essential data is not collected, allowing all city entities, from competent bodies to each citizen, to make smart decisions. Critical data offers convenience, economy, optimal services, and better and more thoughtful design in a sustainable scheme, where problems are not simply solved. Still, based on the stored historical data, their hidden knowledge can reveal trends, allowing relevant agencies to implement preventive policies to avoid complex situations. So, in addition to the direct benefit that any entity of the city may have, these essential data can also provide indirect benefits, as they can be used by academic institutions and research bodies in environmental, social, economic, or transport studies by companies for the appropriate adaptation of their products or services, as well as by non-profit and non-governmental organizations to carry out more effective work [4].

In the face of rising urbanization, city planners are turning to technology to alleviate many challenges in contemporary cities. Smart cities result from deeper technological integration into new or existing urban environments. Building a smart city aims to improve people's quality of life by leveraging technology to improve service efficiency and meet residents' needs. A smart city is a vision for urban development that aims to secure and integrate multiple information and communication technology solutions to manage a city's assets. The smart city is concerned with how the city's "organism" functions as an integrated whole and survives in harsh environments. A city's energy, water [5], transportation, public health and safety, and other aspects as critical infrastructure run smoothly while providing a clean, economical, and safe environment to live and work in [6].

In practice, these transformational impacts will result from the combination of three components of technology: low-cost logic controllers, millions of sensors attached to devices scattered around a city, and a network that links all of these nodes and allows real-time communication. Smart cities rely on networks to ensure the supply and delivery of the functions. Such network connections will allow for more effective and efficient delivery of urban services. Also, these networks aim to present conservation opportunities, improve efficiencies, and, most importantly, enable coordination among city officials, infrastructure operators, public safety officials, and the general public [7].

Many urban areas, including energy supply, transportation systems, and telecommunications infrastructure, have started to include parts of a "smart grid" - or a network of linked sensors inside the city with many advantages. On the other hand, increased connection brings potentially severe cyber security concerns that have yet to be fully identified and managed [8].

Based on the above, information security is critical in intelligent cities for ensuring higher levels of confidentiality, availability, and integrity, as well as the stability required by national services and organizations to support sustainable and livable intelligent environments. Although smart cities are intended to boost productivity and efficiency, they may pose severe hazards to inhabitants and authorities if cyber security is not prioritized.

For example, the growth of the Internet of Things (IoT) reveals new vulnerabilities for intruders and other hostile actors to exploit. There are many possible vulnerabilities and techniques with billions of linked 'things' installed in smart cities throughout the globe [9].

In summarizing, the following are the most significant security challenges for smart city environments [10]:

A large and complex attack surface: As cities become more intelligent, they will incorporate more systems and "systems of systems," increasing the risk and impact of an attack and necessitating better control and visibility. Furthermore, the integration of vendor solutions increases the complexity of intelligent city systems, particularly during rapid technological transformations.

Inadequate oversight and organization: Complex systems will necessitate more robust management and governance capabilities; keeping leadership fully informed of complex occurrences will require additional resources and capabilities.

In the face of a constantly developing cyber threat scenario, the research community must supply threat information, prevention, and reaction as smart cities will give enter-prizes unparalleled economic potential. However, because of the significant increase in interconnected devices, cyber threat actors will be presented with an unprecedented attack surface. Securing smart cities must be a collaborative project involving local administrations and private sector organizations with an immediate stake in the city's stable function. Also, ensuring that smart cities are cyber-safe will need identifying and prioritizing critical assets and behavior-based security - creating a baseline for the routine functioning of vital support. It should be emphasized that a Smart City's smartness is measured by seven aspects in which the city should excel smart government, smart mobility, smart environment, smart living, smart healthcare, smart economy and smart people [1]. These critical assets must guarantee that all sections of the city conform to minimal benchmark, a policy of quick component replacement in case of breach or failure, and a safe segment of critical private assets from the public network. In this sense, this research aims to identify the cyber threats, attacks, and countermeasures based on the seven primary domains of smart cities listed above.

The following is the structure of the study: The next section, Section 2, provides an in-depth description of the primary threat's attacks, and countermeasures that are present in the seven domains of the smart city networks, how they function, and the associated effective solutions that have been proposed in the most recent literature. The primary findings from our investigation are discussed in Section 3, which summarizes our results, makes our conclusions, and discusses potential avenues for further study.

**2. Cyber Threats, Attacks and Countermeasures**

Improving residents' living standards is the primary goal of constructing smart cities. From this point of view, a smart city utilizes its resources more effectively and produces an intelligent ecosystem using the capabilities of digital technology. It requires more inventive urban transport networks, better water supply and waste disposal facilities, and more efficient methods to light and heat buildings in living places. It also involves more creative ways to deliver democracy and decision participation to citizens. It also means having a clean and safe environment with accessible public areas and user-friendly financial services that cater to the population's requirements without restrictions or exclusions. It should be emphasized that this connection is achieved with the optimal use of ICT, which also makes smart citizens of a smart city [2], [8].

This study presents an overview of the most dangerous cyber threats, attacks, and countermeasures to smart city networks based on the seven primary domains of smart cities [11]. In this context, we will examine indicative but very characteristic cases of cyber threats related to these categories and the countermeasures proposed in the recent literature.

*2.1 Smart Government*

Smart governance [12] encompasses services that reflect aspects of political participation and opportunities for citizens' social inclusion in the administration's operation. It also encourages the most efficient administration at the lowest possible cost, in which human resources and available resources are fully utilized, additional training opportunities are provided, and the processing of bureaucratic tasks is promoted digitally, removing the citizen from an unfair waste of time [13]. Furthermore, smart governance empowers citizens to participate in public decision-making and city planning, increasing efficiency and information transparency.

It is necessary to have policy guidelines in place in smart cities to guarantee the smoothness, security, integrity, and secrecy of smart governance [14]. The use of technology in smart governance must comply with a nation's laws to be considered legitimate.

The city's governance services are becoming smarter and more complex, but these users must be prepared to confront cyberspace security using more sophisticated technology and infrastructure. Governance service providers are facing growing problems in adopting a culture of security and information confidentiality as a critical component of their services [15], [16].

These difficulties will endure as long as regulatory oversight, and information security risks exist. Improving the information security culture inside these firms will most certainly aid in the safety of governance information, transactions, and personal information and the continuing performance of essential governance processes. For example, risk identification and mitigation must occur at intersections and in separate domains in the governance system. The designation from the inter-sector dimension considers the linkages of the national or municipal sector and the interdependence of the other domestic and foreign sectors. Furthermore, the intersectoral extent indicates how risk is distributed throughout interrelated systems (contagion risk) because of parallels in concentration risk and connectivity. Meanwhile, the inter-time dimension specifies how hazards in the planned system rise over time, primarily risk caused by procedural behavior shifting from one sector to another [17].

In a Smart Governance scenario, municipal authorities and local administrations strive to structure and arrange city-wide interventions across various IoT systems/applications to build an all-in-one and coordinated IoT ecosystem for the full Smart City. However, the administration of a whole Smart City requires a significant degree of responsibility for the many tasks to be completed, such as managing IoT systems (e.g., credential access, control of sensitive data). In addition, technology is necessary to handle these IoT systems in a coordinated and well-managed manner. The technical background required to carry out know-how is not recognized by City Authorities but by professionals in the field [15].

To successfully manage heterogeneous IoT systems deployed by public and commercial organizations, upper-tier authorities need a multilevel governance solution to distribute and transfer authority and duties across various levels of government (top-down governance).

As a result, municipal governments need ICT-based solutions that offer the elements necessary for the Smart Governance of the Smart City. We have emphasized the following desirable features [13], [14], [17]:

Permission hierarchy model. From upper-tier authorities to local authorities, a multilevel governance system must issue and maintain permits in a hierarchical framework (top-down approach).

Keeping track of duties. Because of the variety of IoT systems and functions supplied in a Smart City, the various authorities must administer a role-based access control system to allocate and monitor duties.

Inclusion. An integrated strategy to accommodating heterogeneity within the same ICT-based solution is required to facilitate data interchange. Third-party API (Application Programming Interface) deployment, sharing of credentials and permissions, and so forth.

Support for new systems. A suitable solution that allows the integration of new IoT systems over time, independent of the technology on which they are based, is necessary for a dynamic Smart Municipal, integrating the new IoT system into the ICT-based solution offered by the city authorities.

Safety and privacy. It is critical to include security methods in the elements outlined above. For example, only authorized and authenticated users may manage permissions using well-tested and robust processes.

Several ICT-based solutions are needed to address these aspects. However, for different entities to cover these features, a specific and common infrastructure that provides the ICT-based tools and mechanisms natively to cover these features is required, ensuring that the same rules and regulations guide IoT systems that belong to a Smart City in terms of Smart Governance [18].

From this point of view, governance systems must be built on acceptable standards such as ISO 27001: 2013, a global standard that defines the requirements for designing and evaluating an information security management system. This is critical for any governance transaction service provider for the transactions as part of a governance process must be assured not only for security and secrecy but also for the ease of continuing to conclude. Other references that are often used for security requirements in technology-based services include [19], [20]:

1. COBIT - The Board of Standards Audit and Information Systems Control (ISACA) publishes the Control of Information and Related Technologies (COBIT), which offers a control framework for corporate governance and IT governance.
2. ISO / IEC 15408 - This assessment criteria was established and is aligned with the national security standards.
3. ITIL (or ISO / IEC 20000 series) - This publication presents a collection of best practices in IT services management (ITSM), focusing on the IT service process and emphasizing the user's important role.
4. ISO/IEC 27001 – This certificate is often the most basic Information Technology Standard and determines which other security standards must also comply by a government organization. For example, if a governance domain offers payment services, it must meet, if not exceed, the Payment Card Industry Data Security Standard (PCI DSS).

This paper [4] shows how proper smart governance and security framework on which built adequate security infrastructure could simplify and speed up the certification process and, simultaneously, reduce the cost of certification.

However, the link between smart governance and data security is quite complex. The combination of information security risks and data challenges, such as data transfer, processing, information management, network security, and equipment access, is significantly challenging. The goal is to improve the smart governance operation environment and the security of big data information by studying and improving relevant laws and regulations, strengthening access control processing, improving system protection levels, strengthening the integration of new technologies, and improving network intrusion detection systems, and the sophisticated mechanisms of identity security verification [21]

A typical example is an authentication on smart governance must be more secure to guarantee its security and privacy, and only authorized users should be able to access the services. A robust authentication mechanism is provided in this research to address the shortcomings of previous approaches. Specifically, a three-factor authentication scheme designed especially for the authentication of e-governance applications. For permission, it exclusively uses lightweight XOR and one-way and perceptual hash operations. Numerous factors such as user identity, password, and biometrics are employed for improved scalability and security. The suggested approach has been verified using the popular Automated Validation of Internet Security Protocols and Applications tool. According to the security study, the system is resistant to numerous threats [20].

In this direction, the authors of this article provide a Smart Data Sharing Architecture for a Smart City Environment based on smart contracts and Blockchains. The suggested approach attempts to protect data exchange among various stakeholders by integrating data access control and audit methods. It employs the Proof-of-Trustworthiness (PoT) consensus technique as a complicated security solution. The proposed consensus technique employs a multidimensional Trust Model to determine the total Trustworthiness of a Smart Service Provider. The suggested solution conforms to fundamental privacy laws and regulations, including the European Union General Data Protection Regulation (GDPR) duties [22].

Also, to protect the privacy of this sensitive information, unique software tools and cryptographic security protocols must be employed using the user's settings. Hardware devices must be put inside the linked system to improve these security settings. Because the government is the critical player in delivering electronic services under its authority, this study has previously presented a citizen-centric multidimensional smart card-based E-Governance system. The authors created a cloud architecture as an interconnected governance model to handle a large volume of sensitive information. The new model's viability is shown via Cloud Banking transactions via a linked governance environment [15].

On the other hand, there are several practical proposals for this avenue, like the authors of this study [16] that propose an architecture for the Smart Governance of heterogeneous IoT solutions inside Smart Cities, which includes aspects such as power distribution, security limitations, and scalability, among others. It is a modified version of the fundamental architectural paradigm of smart government systems. It relies on Digital Objects to represent, store, and interact with physical and digital materials. It also uses open-source technologies to create the architectural concept and manage Digital Objects.

Following the proposed transparent and open adjudication procedure, local officials may assign, for example, the administration of the city's smart lighting system. This is accomplished by defining and creating new Digital Objects as part of the Smart City infrastructure for these new actors: (55555/city/company for the company and 55555/users/manager for the Manager or the company representative) and granting permissions to the Manager to handle all infrastructure related to the smart lighting system. Similarly, the Manager may authorize other members of the institution, such as the IoT system manager (55555/users/iotmanager), to assign tasks to trusted employers.

Adopting this multi-level Smart Governance architecture also addresses one of the most challenging difficulties in any IoT-based solution: naming (devices or IoT settings). Because of the Handle System and its global infrastructure, each infrastructure installed following our concept functions as an independent node in a well-structured network that functions as a Domain Name System (DNS). In this approach, any IoT material and Smart Governance-related element may be accessible from anywhere, including external ecosystems such as other companies, municipal authorities, and Smart Cities.

Remote e-voting is unquestionably the apex of e-governance in a prosperous smart city. Remote e-voting is convenient and gives voters simple access. On the other hand, it makes it simpler for election officials to tally ballots and compile reports. Nonetheless, there has been increased criticism of the security and integrity of remote e-voting systems. Concerns have also been expressed concerning the systems' ability to guard against cyber threats. A functional remote e-voting system must fulfill a set of regulatory requirements. This study [23] examines the design criteria for remote e-voting systems as mentioned in the existing literature. The authors investigate whether the current public infrastructure can support an efficient remote e-voting system that fulfills the design criteria. They discovered that the present technological infrastructure is insufficient to enable efficient remote e-voting systems since the technologies that must be installed to meet design criteria are vulnerable to various cyber assaults.

It is widely accepted, because of the decentralized nature of e-voting systems, ensuring transparency and dependability is difficult. Also, maintaining the privacy, secrecy, and integrity of the votes and voters is equally suspicious. One potential answer is blockchain technology, which has ushered in a new age in the digital world. The immutability of blockchain technology and its decentralized design make it ideal for developing a strong and secure e-voting system. It aids in elections by ensuring the system's authenticity, integrity, transparency, secrecy, and non-repudiation. This article assessed the potential and practicality of blockchain technology for electronic voting. In the present E-voting system, the authors have solved all possible constraints. They created a small-scale example e-voting system as a smart contract using Solidity language, which encompasses everything from hosting the election to certifying voters and counting votes. The study also demonstrates how Zero-Knowledge Proof might aid in developing safe, privacy-preserving E-voting systems [24].

However, implementing blockchain technology in voting systems causes severe delays in transaction execution. The software solution is only available to a small group of professionals, which does not inspire trust among voters. This study [25] suggests a novel strategy based on freely accessible verification of all means and procedures that may generate doubt about the accuracy of the vote count or the secrecy of their results. Specifically, the authors presented an e-voting system in which an audit of all the hardware and software of a server that conducts operations linked to potential misuse is performed to assure confidence. The mechanism guards against illegal influence on votes via bribery or other forms of pressure. It also monitors all employee operations to control the server's functioning from its inception until the election's conclusion. Simultaneously, the number of auditors is not restricted, and the verification method is streamlined by using extra software and hardware. Auditors' responsibilities include copying and comparing files. All e-voting system technological solutions, including audits, are easy to grasp and inexpensive.

*2.2 Smart Mobility*

Smart mobility aims to make transportation systems "smarter." Smart transportation networks, in particular, can better serve the public by improving safety, speed, and reliability to local and international accessibility. The availability of information and communication technologies from modern and sustainable transportation systems is critical for consumers to easily plan their schedules while finding the most economical and fastest routes by using transportation-oriented mobile applications. Driver's passports, license recognition systems, car-parking searching, and prediction are other typical applications in smart mobility facilities.

Traffic safety is a significant concern in crowded cities that harms residents [26]. In this regard, IoT can be more proactive in detecting human errors and reducing traffic accidents. In the literature, for example, particular research gives valuable insights into allowing smart mobility systems. In [27], an IoT-based system was created by combining internet platforms and low-cost antenna technologies. The suggested research in [28] demonstrates the viability of monitoring road safety using developments in IoT. It offers a low-cost IoT framework for assessing the safety of a road network. In addition, [29] presents an overview of the key IoT technologies suggested for smart mobility in smart city scenarios.

Furthermore, IoT-based solutions must be employed for various applications and modes of transportation, such as smart traffic, parking, and mobility, in order to create safer and cleaner streets. For instance, in [30] an intelligent transportation system has been designed to recognize, locate, track, and monitor buses through exchanging information and communication. This system is based on the Internet of Things (IoT), radio frequency identification (RFID), General Packet Radio Services (GPRS), Geographic Information System (GIS), and Global Positioning System (GPS), among other technologies. Another project that is mentioned in [31] aims to construct a prototype for intelligent transportation that makes use of a global positioning system (GPS), near field communication (NFC), and temperature and humidity sensors to monitor automobiles and commuter information and the atmosphere inside buses. [32] Suggests a real-time traffic monitoring system as a solution to the issues associated with traffic management and monitoring. To determine roadway problems, the study used the information gathered from real-time traffic monitoring. [33] Create an intelligent transportation system application that uses Internet of Things platforms, Intel Edison, and Raspberry Pi. It has been proposed that information on traffic conditions may be distributed via standard instant messaging services like WhatsApp.

As the above research works indicate, selecting IoT technology is critical for creating smart mobility systems [34], [35]. However, the security provided by IoT devices is not guaranteed. The devices have a propensity to have little computing power, and their hardware restrictions prevent them from having built-in security mechanisms. This makes the devices susceptible to vulnerabilities [36]–[38]. The IoT devices in a smart city's ecosystem control critical transportation infrastructures, so they need strict secure guarantees. In this point of view, a `Cybersecurity via Determinism' paradigm for the next-generation `Industrial and Tactile Deterministic IoT' is presented in this study [39]. Specifically, in layer 3, there is a new addition of a forwarding sub-layer called deterministic packet switches (D-switches), which are straightforward and safe. This sub-layer supports many deterministic Software-Defined Wide Area Networks (SD-WANs), in addition to three new tools that may be used to improve online safety: access control, rate control, and isolation control. To support D-flows, a Software-Defined Networking (SDN) control plane will set up each D-switch (also known as an FPGA) with several deterministic schedules. The control plane of a software-defined network (SDN) can insert millions of deterministic virtual private networks (DVPNs) into layer 3. This paradigm has a number of advantages, including the following:

1. All congestion, interference, and Distributed Denial-of-Service (DDOS) assaults are eliminated.
2. The size of the buffers in D-switches is cut in half.
3. Delays in end-to-end IoT communications were brought down significantly.
4. The D-switches do not require gigabytes of memory to store large IP routing tables.
5. Hardware support is provided in layer 3 for the US NIST Zero Trust Architecture.
6. Packets within a DVPN can be entirely encrypted using Quantum-Safe encryption, which is resistant to attacks by quantum computers using existing quantum algorithms.
7. By using lengthy Quantum-Safe encryption keys, the likelihood of an undiscovered cyberattack against a DVPN may be arbitrarily tiny.
8. Savings can approach thousands of dollars annually via decreased capital, energy, and operating expenses.

Because of this, the attack surface is substantially smaller, meaning there are fewer packets to assault; thus, the danger posed by this attack is significantly decreased. In addition, using encrypted packets for signals eliminates any possibility that a rogue control-plane packet may get past the Authorization Check [39].

To provide safety and information-related applications on the road, the Vehicular Ad Hoc Network (Vanet) has recently received considerable attention in the smart transportation domain. Vanet delivers an infrastructure where vehicles moving on the road can use communications to report traffic congestion, accidents, and road surface conditions to other vehicles. Although Vanet is an excellent cyber-physical system [40], it has several security and privacy problems, particularly location privacy. In order to be broaded applied, they must be enhanced the Vanets applications to preserve the identity and location privacy of cars. However, since a hostile vehicle cannot be followed using a complete privacy preservation strategy in a cyber security scenario, most users would expect a conditional privacy preservation approach to safeguarding systems. Group signatures may be used to provide conditional privacy preservation. However, the calculation costs are relatively high. Unlikable pseudo-ID approaches may also produce dependent privacy preservation. However, revoking a malicious vehicle would result in a lengthy revocation list. Unfortunately, any proposed method for verified cater evocation does not enable forward unlikability. Forward unlikability is a challenging criterion for Vanet systems. If a car is hacked and turns malevolent, the vehicle's license should be canceled immediately. However, the vehicle's previous communications and positions (from before it was hacked) should be safeguarded and unlikable. To address these issues, the authors of this research study [41] provide a lightweight conditional privacy preservation system that leverages basic hash-chain algorithms to allow accurate identity monitoring by a trusted authority and quick local revocation verification on the road. In Vanet systems, the suggested protocol addresses the needs for location privacy, conditional privacy preservation, and forward unlinkability.

In addition, novel security techniques are presented in this study [42] to enable safe certificate revocation, which is regarded as one of the most demanding design challenges in Vanet networks. In this system, each vehicle that receives a communication from another vehicle validates the sender certificate authenticity. The recipient will evaluate the message to see if the sender has a valid certificate. In contrast, if the sender has an invalid certificate, the recipient will ignore the communication. Furthermore, if the sender does not have credentials, the receiver will report the sender to the Road Side Unit and review the message to see whether it is valid. If the information supplied was accurate, the Road Side Unit will provide the sender with a Valid Certificate. Otherwise, the Road Side Unit will issue an Invalid Certificate and add the vehicle's identity to the Certificate Revocation List. When a misbehaving vehicle with a Valid Certificate is discovered, the RSU replaces the old Valid Certificate with a new Invalid Certificate to indicate that this vehicle should be avoided. This occurs when multiple vehicles report to the Road Side Unit that a particular vehicle has a Valid Certificate and is broadcasting incorrect data.

This research paper [43] presents a catalog of potential solutions, each of which, if put into practice, has the potential to dramatically cut the risk of cyberattacks on the communication systems of connected cars in a vanet infrastructure. Five various degrees of security architectures may be chosen from among the defense options that users can apply. But cyber attackers have a large terrain of assaults and objectives, and the variations between innocuous and those damaging are recorded in terms of the number of people killed and the amount of damage done to transportation infrastructure.

Cryptography, zero-knowledge between communication vehicles, and authentication methods with or without a trusted third party are all necessary breakthroughs; in reality, they have shown to be inadequate [44]–[46]. Especially today, when more and more automated vehicles are being put onto the roads, which need to cohabit with other types of motorized and non-motorized traffic participants efficiently and safely. Autonomous cars do, however, run the risk of traditional cyberattacks on the information and operation of the vehicle, as well as a new breed of attacks surrounding things like ransomware, IoT attacks, and DDoS attacks (connected vehicles drafted into Botnet Armies). Because of their interconnected nature, security risks are associated with the networks to which they are connected. This is true regardless of whether we are talking about the financial networks that process payments, roadside sensor networks, electricity infrastructure, or traffic control features. The authors of this paper propose a method for designing safe and secure mixed traffic systems, including automated vehicles and non-automated road users such as pedestrians, bicyclists, and conventional vehicles. This will allow the authors to model safe and secure cooperating Automated Vehicles and road infrastructure. In addition, this method will enable the authors to design safer and more secure Automated Vehicles and road infrastructure. The applicability of the suggested approach is shown with the help of a typical scenario involving the interaction of an automated vehicle with pedestrians at an intersection that does not have traffic signals [47].

The development of connected vehicles, which produce dynamic data through wireless communications, has made it possible to automate vehicles to operate more effectively. This is especially true in traffic signal control, which serves as the structural foundation for the scheduling of traffic flow. On the other hand, wireless communication channels are susceptible to cyberattacks and may constitute a significant risk to dynamic traffic signal control systems. Attackers might manipulate the usual traffic flow to bring up extreme traffic congestion. The authors of this work use Deep Reinforcement Learning to create an intelligent Sybil attack on a traffic intersection. In this attack, connected vehicles with fake identities are optimally placed to change traffic signal timings by corrupting traffic data. This work aims to highlight and exploit existing vulnerabilities in Traffic Signal Control systems. The findings indicate that this attack causes a sizeable increase in the time it takes for vehicles to complete their journeys and results in catastrophic traffic congestion, mainly if carried out for an extended period. This will lead to several serious issues, including increased fuel consumption and air pollution in cities with a high population density. In the face of such sophisticated assaults, the design assumptions behind present Traffic Signal Control systems have become increasingly suspect [48].

The authors of this research paper [49] establish an assessment methodology for cyber-attacks on autonomous cars using preexisting traffic flow models as the basis for their work. They explore the influence of factors such as the percentage of cyber-attacked cars, the intensity of cyber-attacks, the range of cyber-attacks, and the demand for transportation. In addition, the transportation system's performance is evaluated using a set of four metrics, which include efficiency, safety, emissions, and fuel consumption. The numerical simulations show that as the number of cyber-attacked vehicles and the severity of the cyber-attacks increase, the negative impact on traffic flow gradually becomes more noticeable. This manifests as a reduction in capacity, an increase in the risk of rear-end collisions, an increase in air pollutants and fuel consumption, and so on. In addition, if cyberattacks are carried out on location rather than speed, this might result in accidents and decrease traffic operations' efficiency. Therefore, the position-attacked traffic system has a higher overall energy consumption and a higher overall pollution output. The results of this research give helpful information that may be used for projecting future traffic targeted by cyberattacks, conducting complete evaluations of transportation systems, and managing automated highway systems from the point of view of network security.

But as is easily understood, the big problem in smart mobility is the cyber issues of mass transportation in which many people are carried within a single-vehicle. Specifically, the most serious difficulty is the cyber security of public transit [50], [51]. Trains, for example, are the most often used mode of transportation in a modern city, with millions of daily passengers, as opposed to car transportation. Because of system and infrastructure digitalization, automation of railway processes, mass transit concerns, and expanding linkages with external and multimodal systems, the railway industry is experiencing a significant change in its operations, procedures, and infrastructure. Cybercriminals may choose to target ticket vending machines, passenger information screens, and passenger Wi-Fi infrastructure. These systems are becoming more vulnerable to cyber-attacks as they transition from bespoke stand-alone systems to open-platform, standardized equipment built with commercial off-the-shelf components and increased use of networked control and automation systems accessible remotely via public and private networks. Many signals transmitted and received over insecure communication links are critical to railway operations. Constant monitoring, immediate notice of any departure from ordinary circumstances, and decisive measures to resolve the issue allow for effective risk reduction and business continuity. Furthermore, uninterrupted and safe traffic operation depends on recognizing and addressing threats to telecom, train management, and signaling systems as soon as possible. Complete visibility and the capacity to identify and mitigate hazards as they emerge give a route to reducing uncertainty and operational interruptions [52], [53].

Though various studies on critical infrastructures have been conducted from the aspect of cyber security, there has been little research conducted from the standpoint of cyber-physical security in application areas such as railway infrastructure. This is the first complete empirical evaluation of the cyber-physical vulnerability of communication-based train control systems [54]. The writers carefully analyze communication-based train control and the cyber-physical vulnerabilities that may seize control of the train. They discover that a man-in-the-middle assault combined with knowledge of railroad signaling may result in substantial train crashes. They propose a countermeasure for communication-based train control resilience to overcome the issue and meet these difficulties. The primary idea behind this countermeasure design is to create a subsystem with a host that the attacker cannot reach. The cable links the subsystem to the software-defined networking (SDN) switch. The SDN controller, on the other hand, logically disconnects the link between the SDN switch and the subsystem. The subsystem stays undiscovered due to the logical separation while the attacker seeks victims.

Consequently, during a man-in-the-middle attack, the subsystem may go undiscovered. For the subsystem to warn the automated railway protection system of the attack scenario, the SDN switch should detect ARP spoofing. The SDN switch continuously monitors ARP messages and creates an IP-MAC list for ARP messages across the structure's SDN switches. They validate their findings by developing a realistic communication-based train control testbed environment that yields promising outcomes.

In addition, this research [55] looks at the security of railway control equipment against cyber or physical attacks. The authors offer a cyber-physical authentication approach that integrates an add-on security module with a cyber security protocol for cyber-physical device authentication and data transmission. An add-on security module that acts as a 'hot swappable' bump-in-the-wire solution for current control devices and adds extra cryptographic capabilities was offered. Meanwhile, the suggested system offered tamper-resistance security by encapsulating the hardware in a tampered-detection box and employing tamper-resistant hardware to secure the cryptographic key. The device authentication protocol is intended to provide secure communication between the Server and the Control Device. The protocol is divided into four stages: setup, first handshake, acknowledgment/critical exchange, and data transmission. Due to the necessity of storage, RSA signatures were used for the first handshake, and AES with a 256-bit session key was utilized for the data transmission phase. The results of the tests reveal that the overall overhead has a minor influence on the functioning of the Control Device in the railway system.

The criteria for the railway management system are outlined in detail in the work [56], which also provides a summary of the requirements that must be met for the Smart City concept. The authors offer a set of actions for enhancing cyber security and the information power of railway management systems, with the instruments of risk engineering and the knowledge gained from the area of information technologies serving as the foundation for their recommendations. Also, in this paper [57], the authors provide the bases for a combined Safety and Security risk assessment and analysis approach. This approach reconciles the risk analysis processes that are used in both Safety and Security by making relevant connections at different stages of the two processes and by adding cross-cutting steps that are common to both Safety and Cybersecurity.

*2.3 Smart Environment*

A smart environment can make a significant contribution to the development of a sustainable society. The smart environment combines appealing natural conditions such as climate, green spaces, and so on with techniques for limited contamination, optimal resource management, and environmental actions. It is also related to access to services that improve the city's quality of life and the facilities of public spaces on a broader scale. In addition, it is associated with the city's cleanliness, the initiatives that give life and movement, and strengthening security in areas such as local forests, lakes, etc [58]. A smart city can monitor energy consumption, air quality, building structural reliability, traffic congestion, and address pollution or waste using technical management tools. Thus, the sustainable and smart city considers using and producing green and renewable energies, more sustainable food production techniques, or the application of innovative technology to improve resource management (energy, air, water, waste, etc.). Novel environmental Wireless Sensor Networks (WSN) have the potential to monitor the natural environment and potentially anticipate and detect natural disasters [59], [60].

WSN are networks of autonomous sensing devices that monitor physical or environmental factors such as temperature, pressure, sound, vibration, motion, or pollution at several places. WSNs are multi-hop ad hoc self-organizing networks in which all nodes interact wirelessly and use multiple routing protocols [61]. It works in circumstances with limited bandwidth and performance. It is scalable and may accept more nodes or devices at any moment. It is also adaptable, allowing for physical divisions, and all nodes may be accessible through a centralized monitoring system [62]. Because it is wireless, it may be used on a big scale and in a wide range of environmental applications or sectors. Furthermore, it employs multiple security methods based on the underlying wireless technology, resulting in a dependable network for specific users [63].

Smart cities can save energy and reduce carbon issues using WSN. Specifically, through a large number of wireless sensors, can collect environmental factors and then return the information to the backend monitoring server. This study [64] presents a WSN using environmental sensors and controllers to adjust the energy consumption of electrical appliances. Because the plethora of wireless nodes is exposed to physical or logical access in remote urban areas, the authors performed a massive simulation of DoS attacks or external damage by human manipulations. The authors use a sophisticated analysis of various packet loss patterns to identify the potential damage and examine the abnormality. After identifying damage by the logical or physical attack, each node is used by the different queue management models to collect environmental data.

It is difficult to approach or work at the WSN since it operates in a complicated setting. Because nodes are open, they are susceptible to numerous assaults. Traditional security systems will also mistake nodes deployed in a complex environment with poor-quality connections or a poorer condition (less energy or a higher workload) for malicious nodes. In WSN, the trust and reputation model may be employed to mitigate the harm caused by malicious nodes. However, trust and reputation models have a sizeable false-positive rate since a node with less reputation is evaluated undesirable owing to the communication context. This study [65] provides trust and reputation-based harmful node detection techniques with environmental factors to prevent malicious nodes from interfering with or selectively forwarding attack nodes. Environmental parameters may be solved using machine learning's linear regression and combining node energy, data volume, number of nearby nodes, node sparsity, and other deterministic characteristics. The environmental parameters are then used to estimate benchmark trust based on the ecological factors. The Gaussian radial basis function is simplified to compute the similarity between the benchmark trust sequence and the cycle reputation sequence. Furthermore, environmental settings provide three reputation intervals and an adoption threshold span to detect malicious nodes based on the work environment and node statuses. The simulated findings demonstrate that environmental factors increase the detection of malicious nodes by more than 1% compared to comparative methods and minimize false positives by more than 1%.

To communicate outside of the wireless communication zone, WSN employs mobile nodes. Wireless ad hoc network routing protocol attacks degrade network performance and dependability. Malicious nodes advertise the shortest route between source and destination in active black hole attacks on wireless networks, resulting in routing table alterations and packet loss. Self-security management is a hot topic in WSN-related environmental applications. For example, this paper [66] provides the Grouped Black Hole Attack Security Model (GBHASM), which prevents grouped hostile nodes from advertising the shortest route between source and destination, preventing routing table alterations and packet loss. The GBHASM proposed is separated into two components. The first module describes how a new node will join the network, while the second handles communication. The joining request is received from the new joining node in the replay. It sends membership acknowledgment and waits for replication approval. If the request is not accepted within a specific time frame, it will be rejected; if approved, the demand for its details will be delivered. The same procedures will be repeated till the process is completed. Information received from a new joining node is recorded in the database, given a new node code, and the updated node code table is propagated on the network. The second model deals with network communication activities. After joining the network, the node sends out queries for the quickest network route. Every node will compare the node code, and if the key matches within a specific time frame, information about the packet will be revealed. Otherwise, the time-to-live packet will be sent on to the next node.

Although prevention and monitoring techniques may lower the risk of cyber assaults, the residual risk in vital infrastructures or services might still be unacceptable. Resilience, or a system's capacity to survive evil occurrences while retaining adequate operation, is a crucial attribute of such systems. While numerous resilience indicators have previously been proposed, there is limited experimental data on the cyber security of CPSs. This study [67] aims to provide a model-free, quantitative, and general-purpose assessment approach for extracting resilience indices from data sources such as system logs and process logs. The authors evaluate four resilience indices from a broad range using an actual wastewater treatment plant model and modeling assaults that interfere with a vital feedback control loop. The findings reveal that although the selected indexes varied in their behavior and susceptibility to certain assaults, they can all summarize and extract valuable information from large system logs. The proposed method includes deriving performance indicators from observable data that does not need system dynamics understanding.

Waste management and recycling are critical to remaining sustainable and clean in contemporary metropolitan settings. Solid waste management, disposal, and recycling are all difficulties in many major cities across the globe. Combining IoTs with deep learning provides a modular approach for data classification and real-time analysis. This article [68] depicts an effective smart waste management and classification system based on IoT and deep learning. The study proposes an architectural concept for a microchip-based rubbish bin that connects with the approach to collecting waste as rapidly as feasible. The Internet of Things (IoT) is employed in the recommended data monitoring system to provide real-time data control. This smart trash management and categorization strategy also included a waste classification algorithm based on convolutional neural networks. This trash classification approach will be utilized at the waste-collection facility to separate garbage into several categories to enhance recycling. This suggested system provides comprehensive garbage management and recycling solutions for smart cities, including waste collection, management, and categorization.

Modern technologies enable Water Distribution Systems (WDS) to provide improved water supply, storage, distribution, and recycling services. They help with real-time monitoring, automation, and management. However, the limitations of these technologies expose the WDS to cyber-physical threats. The primary aim of cyber-physical assaults is to interrupt regular operations and tamper with crucial data, which negatively influences the WDS. As a result, it is critical to design and deploys solutions to improve WDS security by detecting and mitigating cyber-physical assaults. The authors of this research [69] thoroughly investigate typical cyber-physical assaults and common detection strategies for the WDS. They contrast assaults and detection approaches, focusing on concepts, methodologies, evaluation findings, benefits, limits, etc.

The increasing number of successful and attempted attacks on critical infrastructures, such as power grids and water treatment plants, has resulted in an urgent need to develop and implement methods for detecting such attacks, which are frequently carried out by state actors or insiders in the targeted organization. This research [70] aims to provide a case study of an infected wastewater treatment plant (WTP) that used a live memory dump acquisition Imager. The forensic carving procedure is removed in bulk, and features are extracted from the memory dump. In addition, this study [71] emphasizes one method that tries to identify assaults that compromise one or more actuators and sensors in a plant by successfully infiltrating the plant's communication network or accessing the plant computers directly. This technique, known as Distributed Attack Detection (DAD), may detect assaults in real-time by spotting irregularities in the behavior of the physical process in the plant. The use of monitors that are actual implementations of the invariants obtained from the plant architecture is how anomalies are discovered. Each invariant must remain valid during the whole plant operation or while the plant is in a specific condition. A functioning water treatment facility was used in an experiment to evaluate the efficiency of DAD, and it was proven to be successful at detecting sneaky and coordinated assaults. Also, this study [72] aims to provide a technique for enhancing operational security in a wastewater treatment plant and to demonstrate how this approach can be used in a particular setting.

In parallel, this study [73] suggests a cyber-security monitoring system that connects time-series event data, visually [74] depicting security occurrences. It provides a predictive prediction of probable circumstances based on established situations. The motivation for this stems from the requirement to comprehensively understand security events or attacks on a network and information about the intensity and propagation pattern. Furthermore, it may assist in business choices by identifying or comprehending the link between computer equipment and their business/information technology services.

In recent years, the smart energy grid has steadily become the usual development trend in the world's power business. It is an improvement to the old system that incorporates technology and communication into the present grid. This results in a more efficient grid that decreases energy demand peaks and can efficiently integrate renewable resources (at naturally varying levels) into its network. To increase their security, smart grids have included physical control, data encryption, and authentication technologies. However, there is still a shortage of timely and efficient detection tools to keep the grid safe from unwanted breaches. In response to this issue, a machine learning-based methodology for detecting smart grid DoS assaults has been developed in this study [75]. The model initially gathers network data, picks feature, uses PCA to reduce data dimensionality, and then employs the SVM algorithm to identify abnormalities. The study exploits the attack vulnerabilities of smart meters and data servers to add data collection and intrusion detection modules between smart meters and data servers, aiming at the design structure of the smart grid. In the smart grid, real-time data capture and detection are possible. When DoS attack activity is recognized, the alarm system is initiated to handle the alert.

Cyber security must be a primary priority for electric power providers installing smart meters and smart grid technologies. Despite the well-known benefits of smart meters, how and to what degree cyber assaults might disrupt smart meter functioning and remote data collecting about power use from client locations is unclear. To answer these issues, this study [76] tested a commercial-grade smart meter in a controlled lab and assessed its operational integrity under cyber-attack situations. In addition, the False Data Injection Attack (FDIA) is a way of disrupting the security of the power system based on meter measurement. FDIA detection researchers are now focused on detecting its existence. FDIA location information is also critical for power system security. Finally, in this study [77], identifying the meter's FDIA is seen as a multi-label classification task. Each label denotes the current condition of the respective meter. The multi-label decision tree approach is used as the classifier in the ensemble model to discover the precise position of the FDIA. This approach does not need power topology information or statistical knowledge assumptions. The suggested method's performance is validated by numerical tests using the IEEE-14 bus system.

*2.4 Smart Lining*

Smart living is intended to optimize and manage facilities. It emphasizes one of the primary goals of the sustainable and smart territory, which is to enhance the quality of life of its residents. The Smart Living axis is built on three key pillars: civic safety, social cohesion, and tourist attraction. It is a solution that maximizes the city's infrastructure while improving people's quality of life. Allows for real-time control, forecasting, and optimal asset optimization and management. The following are some smart living applications [78]:
1. Detectors of fire: It is feasible to monitor, detect, and prevent fires in the urban environment with this 24-hour program. Furthermore, the detection methods may approach the various sources of fire efficiently to manage the fire more effectively.
2. Intelligent video surveillance: to improve public safety and to use predictive analytics to optimize traffic flow and citizen safety.
3. Sports facility management: Smart living solutions include controlling capacity and performing centralized administration of sports facilities. It is also beneficial for making judgments based on historical data, identifying deficiencies or requirements, and implementing management and infrastructure upgrades.
4. Smart home automation: It monitors and regulates home smart applications such as temperature, humidity, electric equipment, security, and so on in real-time.

Smart living ICT utilities allow operations to decrease resource overconsumption, such as water and gas, while increasing economic development and environmental protection. But the linked smart house poses a variety of security risks. To begin with, individual smart living gadgets may not be secure. Some IoT home devices are hurried to market, and their security may be compromised. In certain circumstances, user manuals fail to address privacy issues or provide sufficient information to ensure the device's security. Baby monitors and security cameras, for example, have been hacked, enabling hackers to view inside a home [79].

Second, the home network may not be secure, and any data stored on it may be available to intruders. An intruder, for example, may follow the device use habits to determine when users are out from home. If the main internet account controls the home network, data from IoT devices may also be in danger. Any flaw might expose personal or private information, including emails, social media accounts, and bank credentials. It is essential to ensure that a single insecure IoT device does not jeopardize the home network's security. Many customers operate their linked house through smartphones, making it a valuable database for anybody looking to hack into their life [80].

As follows from the above, it is critical to describe and comprehend the direction and development required to guarantee that, as Smart Home Systems become more prevalent, the security and functionality of these systems are maintained. Form this spirit, the goal of this study [81] is to identify the hazards associated with smart home systems and research ways to reduce such risks. Also, it provides a comprehensive analysis of the techniques currently used by intruders, the reasons for adopting these methods, and what might be done differently to enhance smart home security. Also, this paper [82] presents and discusses the threats that can affect smart living systems and define the requirement to improve secure communication between the smart home devices and applications that remotely control the home devices.

Vulnerabilities in IoT-based systems provide security risks and obstacles for smart applications. One of the most significant impediments to IoT-based systems has been identified by the low-level security provided. The smart home environment presents novel security, authentication, access control, and privacy concerns due to its internet-connected, dynamic, and diverse nature. The IoT-based smart environment requires an attack model and a risk management framework to improve information security and integrity. This research [83] provides a finite state automata-based attack model for investigating smart home-based security assaults and assessing their effect using our suggested risk management framework for mitigating IoT smart home-related attacks. An examination of typical attack behavior and the risk management framework demonstrates that the proposed approach is feasible and effective and can be used in many smart home applications.

IoT implementation in the Smart home sector is complicated since the devices utilized in such platforms vary in size and computing capacity. The capacity to impose security on such machines depends on how well the authentication procedures are carried out. Against this backdrop, this paper [84] is designed to thoroughly examine possible authentication risks and attacks on IoT, specifically in the Smart Home sector. The significant concepts offered in this study on potential authentication risks and assaults on IoT in Smart home applications are primarily influenced by a careful literature assessment of relevant work in IoT.

This research looked at the system architecture of a smart home ecosystem, vulnerabilities, potential assaults, requirements, and post-attack settings [85]. Proposes an Architecture Analysis and Design Language tool to model the smart home architecture, which was then visualized using a complex graph tool against a security policy. The attack graph highlighted the need to address security considerations while creating smart home systems and identifying potential threat landscapes.

Based on the notion that the primary goal of home automation is to control home devices from a single place. This article [86] presents the design and implementation of a safe framework that provides flexibility and security for smart home systems based on CPS and IoT. As a result, the authors suggest a safe design that protects the system from external internet infections. They address home automation security concerns by installing a secure firewall software solution. A secure firewall identifies and warns the user of specific security vulnerabilities before launching its mitigation technique. Internal security additionally offers encryption for communication and protects the home automation system from unethical acts. This technology (cipher firewall) determines the user privacy problem since it does not allow external threats. Users may monitor smart home activity by attaching static IP addresses to the system. Users may connect to a home coordinator already tied to home automation through cellular internet. Users may turn on/off their TV, door, lights, heat radiator, air conditioning, and water appliances using their connection.

The comparatively inadequate information security of smart home system (SHS) devices may jeopardize consumers' privacy. The authors of this paper [87] propose a novel block data structure based on homomorphic encryption (HEBDS) to record the SHS device information transaction. This study presents a homomorphic consortium blockchain based on the standard smart home system for SHS-sensitive data privacy preservation (HCB-SDPP). To validate operational nodes and transactions in SHS, the authors introduce verification services constituted of verification nodes to our model. They create an encrypted algorithm based on Parlier encrypted for privacy protection using the HCB-SDPP paradigm. To validate the HCB-SDPP architecture, they encrypt sensitive data from all gateway peers and submit it to the consortium blockchain. Following homomorphic encryption processing, they assess the security of sensitive data. The authors also develop attack trials in the experiment to target various peers on the consortium blockchain under the HCB-SDPP architecture. The impact on the whole model is examined if these nodes are vulnerable. The simulation results suggest that the HCB-SDPP model is more successful than SHS at protecting client privacy.

Wireless home alarm systems are becoming more popular, but their security has received little attention. Existing attacks on wireless home alarm systems use networking protocol flaws while ignoring issues caused by the physical components of IoT devices. The authors of this study [88] demonstrate novel event-elimination and event-spoofing attacks against commercial wireless home alarm systems by interfering with the reed switch in practically all COTS alarm sensors in this research. In both assaults, the external adversary controls the state of the reed switch with his magnet to either delete valid alerts or spoof false warnings. Also, demonstrate a novel battery-depletion assault using programmed electromagnets to fast and quietly drain the alarm sensor's battery, intended to last a few years. Extensive tests on a sample Ring alarm system indicate the effectiveness of these assaults.

On the other hand, several inspired smart living applications fulfill the privacy and security standards as [79] The primary purpose of this proposed project is to use new IoT technologies to enable the senior population to self-manage their health and remain active, healthy, and independent for as long as feasible in a smart and safe living environment. An open-source, comprehensive IoT ecosystem is proposed. It includes the following processes: data gathering, data transportation, data integration, processing, manipulation, and computing, visualization, data intelligence and exploitation, data sharing, and data storage. This unique cloud-based IoT ecosystem serves as a one-stop-shop for integrated smart IoT-enabled services to assist elderly persons (65 and older) who live alone at home (or in care homes). Another breakthrough is this system's design and implementation of an integrated IoT gateway for wellness wearable and home automation system sensors with diverse connection protocols. The smart living system and services address smart health and care, smart quality of life, and the social community. The system is developed using the user-centered design process to enable active user interaction throughout the project lifecycle and relevant standards and compliances (e.g., security, trust, and privacy) that are followed to increase user-friendly adoption.

Access-control rules in smart buildings are becoming more dependent on context, such as who is taking action, if there is an emergency, or whether an adult is around. The extensive literature on context sensing might be used to provide contextual access control, but it mostly overlooks threats, adversaries, and privacy. The authors of this work [89] reassess the literature on home context sensing from the standpoints of security and confidentiality. They describe a unique threat model in smart homes focusing on non-technical adversaries' capabilities. In this model, replay, mimicry, and shoulder-surfing assaults are significantly more common. They also synthesize circumstances pertinent to home access control, matching them to existing sensors. They then organize the sensing literature to provide a decision framework for home context sensing that considers security, privacy, and usability. Using their approach, they discover that present sensors do not adequately reduce potential hazards in houses. Some sensors are vulnerable to primary threats such as physical denial-of-service attacks, making it simple to circumvent restrictions based on the lack of a characteristic. Many sensors capture more data than necessary and are useless for all user groups or scenarios.

Simultaneous advances in the Internet of Things and Machine Learning have resulted in exciting multidisciplinary applications, such as classification tasks based on data provided by smart devices for different applications such as resource allocation, security, and activity categorization. However, such applications may be vulnerable to adversarial scenarios. The authors of this research [90] create a white-box adversarial attack technique to produce adversarial instances for data acquired from smart meters placed in residential homes and show that their statistical features are indistinguishable from actual data points. Adversarial machine learning, a method that uses false data to trick algorithms, is a developing danger in AI and machine learning research. The most typical reason is to cause a machine learning model to malfunction [91]. An adversarial attack might include training a model with erroneous or misleading data or injecting deliberately crafted data to confuse an already trained model. The attack technique focuses primarily on Deep Learning-based models used in smart home device categorization. Because the adversarial data points are statistically indistinguishable from the actual data points, non-machine Learning-based solutions may be unable to address the issue given by hostile instances. The suggested strategies' efficacy is proved using the publicly accessible United Kingdom-Domestic Appliance-Level Electricity smart-meter dataset.

*2.5 Smart Healthcare*

Quality, results, and value are the outcomes of technological advancement in the health industry. Patients need excellent and personalized services. Thus, it is critical to invest in this area. Digital healthcare is a movement that entails precisely leveraging new technology to increase the degree of support while keeping costs as low as feasible. Smart healthcare entails not only the adoption of new goods and technology for diagnosis and treatment but also a more considerable interchange of information across parties, a more active role for patients during treatment, and improved clinical data management [92], [93]. Human and non-human players in smart healthcare include physicians, patients, hospitals, and research organizations. Smart Healthcare strives to make patient care safer, better, and more manageable. IoT connection is transforming healthcare by linking patients and healthcare professionals in novel ways [94]. Wearables, skin sensors, home monitoring tools, and other IoT-connected gadgets allow deeper medical insights into symptoms and health patterns, new degrees of remote care, and more control over patients' care and treatment. Sensors, like those used in other IoT-connected applications, are critical in allowing the gathering and analysis of real-time patient data. Artificial Intelligence (AI), the Internet of Things (IoT), the Medical Internet of Things (MIoT), edge computing, cloud computing, big data, and next-generation wireless communication technology are at its heart [95]. This allows healthcare practitioners to spend more time with patients and treating diseases and less on logistics.

Smart health is an important field that continually analyzes patients' health, alternative treatments, and cutting-edge disease-fighting technology. The purpose of smart health is to provide individuals with medical services at any time and from any place. Smart health monitoring devices are often connected through wireless networks, which are vulnerable to cyber threats. However, various risks may endanger these health monitoring applications and systems [96], [97]. These include Denial of Service (DoS) attacks, Fingerprint and timing-based eavesdropping, router attacks, select and forwarding attacks, sensor attacks, and replay attacks [98], [99]. In this paper [100], the authors investigate the consequences of these attacks on health monitoring systems and recommend some interventions based on their research findings.

The most challenging aspect of smart health apps is protecting data from numerous assaults while using simple approaches and algorithms. Given the security problems associated with implementing smart health apps as the primary data collecting and transmission source, cyber threats are categorized to create a viable defense strategy. This research [8] contributes to the analysis of security and privacy in the context of smart cities for healthcare applications in two ways. On the one hand, an overview of several IoT applications and their cyber vulnerabilities is provided. On the other hand, a complete assessment of potential solutions to the issue of cyber assaults is given.

In recent years, smart healthcare has gained popularity. Because of the data, it collects, a more secure mechanism is required to maintain security and privacy. As a result, these techniques may also guarantee security and privacy in Smart health. The goal of this review paper is to assess individuals' concerns with security issues in their smart homes and to start a debate about how healthcare equipment that comes packaged with future houses is susceptible to cyber-attacks that result in data breaches. Blockchain is proposed in this study [101] to increase security and privacy in smart healthcare. Using this platform, the writers discussed the emerging blockchain technology and how its components may help healthcare flourish while maintaining total security. The proposed platform uses blockchain technology and considers the client control over the healthcare data, implying the ability to share data with a specific association or person on a case-to-case and field basis.

Humans benefit from the fast progress of the MIoT connected with biosensors in various ways, including smart healthcare systems (SHS). The combination of MIoT devices and the increasingly networked nature of the healthcare environment enables healthcare providers to provide more efficient and effective emergency and preventative medical services to their patients. SHS offers several chances for healthcare professionals and institutions to monitor patient's health remotely. The health statistics acquired by SHS is kind. However, SHS exposes patients' health data to various assaults. The most complicated difficulties in smart healthcare systems are ensuring patient health data confidentiality and privacy. This paper [102] explores health data confidentiality, MIoT healthcare security concerns, and the rules and regulations involved in establishing a smart healthcare system. It also describes the many prevalent principles and assaults exposed to corporate digital assets and patients' health information, as well as the necessary solutions for overcoming the present obstacles, such as a cryptographic function and a communication protocol.

For example, this work [84] provides a complete overview of possible authentication risks and attacks on IoT healthcare devices in the Smart healthcare area. Furthermore, utilizing a Hybrid cryptographic technique, the authors of [103] established a secure cloud storage solution for healthcare data. A symmetric algorithm encrypts data, while an asymmetric algorithm encrypts keys. The performance and security of the proposed approach were measured and compared to a well-known current technology. Because it is based on a modular exponentiation process, the RSA method often performs poorly. As a result, the authors used the Montgomery modular multiplication technique to enhance the RSA implementation. Blowfish encryption is used when storing health-related data in the cloud, and keys are handled using the improved RSA technique. This hybrid technique provided advantages such as quick encryption, large prime numbers for essential creation, and efficient key management. The simulation results reveal that the suggested hybrid technique's encryption and decryption time is faster than other approaches examined for comparison.

Important points to discuss when exchanging information in a smart health system dealing with critical patient data are the technique to communicate across institutions and safeguard patients' private data. This work [104] proposes a method for saving image-type data using visual cryptography and distributing the data utilizing secretive sharing using practitioners' passwords. However, if penetrated, the underlying infrastructure might result in private data leaks and the destruction of healthcare records dependent on the control commands supplied by the attacker. The authors of this paper [105] concentrate on several degrees of security associated with the storage and transmission of healthcare information. Furthermore, they test some of the suggested approach's relevant characteristics using the Access Control Policy Testing tool to establish its practicality and examine the state-of-the-art in the subject. But, because the present record management system cannot full handle privacy and integrity, the health sectors nowadays using blockchain technology to store health data more safe, confident, and decentralized manner. This study [106] offers a blockchain architecture for Electronic Health Records to safeguard private data based on the Elliptic Curve Cryptography Algorithm. In parallel, this study [107] proposes a realistic solution based on the unique characteristics of blockchain, where the distributed ledger technology is thought to be unbackable. The authors created a blockchain model to safeguard data security and privacy, assure data provenance, and give patients total control over their health information using the smart contract feature, a programmable self-executing protocol operating on a blockchain. This concept delivers a patient-centric procedure by customizing data segmentation and creating permitted lists for physicians to access their data. It assesses the model's feasibility, stability, security, and robustness. In addition, this article [108] offers a permissioned Ethereum blockchain that connects hospitals and patients all over the globe. The proposed system employs a mix of symmetric and asymmetric key encryption to enable safe storage and selective access to records. It provides patients total control over their health information and allows them to grant or deny access to their records to a hospital. The authors stored records using IPFS (interplanetary file system), which has the benefit of being dispersed and assuring record immutability. The suggested methodology also keeps illness data without invading any patient's privacy.

The rising availability of healthcare data necessitates precise analysis of illness diagnosis, progression, and real-time monitoring to enhance patients' therapies. Machine Learning (ML) models are used in this context to extract significant characteristics and insights from high-dimensional and heterogeneous healthcare data to identify various illnesses and patient behaviors in a Smart Healthcare System (SHS). However, recent studies reveal that ML models employed in different application areas are susceptible to adversarial assaults. This work [109] describes a novel adversarial method for exploiting the ML classifiers used in a SHS. An attacker with a rudimentary understanding of data distribution, the SHS model, and an ML algorithm may launch both targeted and untargeted assaults. Their attack employs five different adversarial ML algorithms to carry out various malicious behaviors on a SHS (e.g., data poisoning, misclassifying outputs, and so on). Using these adversarial capabilities, the authors modify medical device readings resulting from the SHS to change patient status (disease-affected, normal condition, activities, etc.). Furthermore, we undertake white and black box attacks on a SHS according to an adversary's training and testing phase capabilities. They also assess the effectiveness of their work in various SHS settings and medical equipment. Their rigorous study demonstrates that the suggested adversarial approach may severely reduce the effectiveness of an ML-based SHS in properly recognizing illnesses and normal patient behaviors, resulting in incorrect treatment.

*2.6 Smart Economy*

A sustainable and smart city is a fertile platform for innovation and new business models [110]. This vision is supported by ideas such as sustainable entrepreneurship and the circular economy. These innovative approaches promote local and global financial ecosystem linkages while fostering long-term economic competitiveness. The concept of smart economy is a prosperous economic prototype built on technology innovation, resource efficiency, sustainability, and high social welfare. It encourages innovation and new entrepreneurial activities while increasing productivity and competitiveness with the overarching objective of enhancing inhabitants' quality of life [111]. This technology-driven, interconnected system employs ICT applications for economic progress, urban planning, and public health improvement. It combines enhanced creativity with improved production, efficiency, and competitiveness. It is distinguished by several new flexible kinds of labor and start-ups. A smart economy is projected to produce more goods and services while using less energy and emitting less pollution, as well as offer social benefits [112]. For example, in Jakarta, as a smart city, the smart economy idea being applied aims to encourage an entrepreneurial and innovative spirit in the society to attain high productivity. Jakarta also has several Smart Economy projects [113], [114] that benefit its inhabitants. Among these programs are [115]–[117]:

1. JakPreneur: Jakarta's MSME development initiative encourages the creation and collaboration of an entrepreneurial ecosystem. MSME actors will get training, coaching, marketing, and even instruction on obtaining funding via JakPreneur.
2. JakPangan: a feature that provides information on the cost of Jakarta's primary commodities.
3. JakNaker Platform: a job-search portal designed to make it simpler for Jakarta citizens to find jobs.
4. JakOne Pay: a non-cash payment option developed in partnership with Indonesia Bank [12].
5. JakLingko Card: The JakLingko card may be used to pay for many kinds of transportation (buses, microbuses, railways, etc.).

To attain the degree of smartness that cities manage, smart cities are often tied to development initiatives. Several studies look at intelligent smart city performance utilizing various approaches and assessment criteria. Although there is no way to do such an assessment, we have found that the Smart Economy is essential in assessing smart cities. The smart economy was a significant feature in determining smart cities in 25 of the 30 studies we examined between 2015 and 2020. The authors of this research [110] underline the relevance of the Smart Economy component in smart city development and demonstrate how it is assessed and evaluated in various evaluation methods. Furthermore, it delves into smart economy-related metrics utilized in multiple evaluation methods. The authors use a theoretical approach combined with quantitative and qualitative data to investigate similarities and relationships among the best-performing cities in the smart economy.

In order to accomplish urban management services, business survival growth, and inhabitants' productive lifestyle, a smart city is the application of a new generation of information technology. The smart economy boosts core competitiveness and may give citizens various economic options. Using China as an example, this paper [118] proposes a smart city development path based on the current state of smart city construction: use Internet technology to build big public data, strengthen the structure of digital infrastructure projects, ultimately mine data, lay a good technical foundation for smart city construction, integrate urban development with the Internet, and gradually promote harmonious town development. Moreover, this study [119] proposes a new service-oriented manufacturing business model called a smart linked product-service system (PSS) against the rapid rise of customized service demand, the digitalization of goods and services, and the socializing of service resources. The issue of CPS-oriented smart connected products and associated service resources is explored from the service flow and product serviceability, as well as the smart corresponding product service system design and socialized service crowdsourcing setup. It is divided into three sections: analysis of personalized service demands using service flow design and product minimum service capacity modeling method to meet service requirements; CPS embedded Web services modular design method for smart connected product; and community construction of service resources using federated learning and service order-driven social crowdsourcing configuration. The objective is to create a new generation of service-oriented manufacturing models that are smart, connected, efficient, and adaptable.

Growing a local economy based on sharing data from IoT devices and other open data that can be utilized in apps to better the lives of its residents is one way a city might become smarter. The authors of the [120] investigate how blockchain and other distributed ledger technologies may be used to build a decentralized data marketplace. They explore the potential advantages of such a decentralized architecture, define several features that such a decentralized marketplace should include, and demonstrate how they may be incorporated into a holistic system. They also describe a basic, smart contract implementation of a decentralized registry where data owners may publish items for prospective purchasers to retrieve.

There are several hurdles to overcome to advance this decentralized marketplace. These are some examples:

Managing system complexity: With so many moving parts and components, the decentralized marketplace system may become fragmented or difficult to scale; how can this be avoided? On the other side, maintaining a tight foundation may hinder scalability.

Economic incentives and centralization: Without sufficient financial incentives for all parties involved, the decentralized marketplace may fail to operate successfully. On the other side, it may be required to carefully construct the incentives so that additional decision-making power and data do not get concentrated in the hands of specific parties, thereby distorting search, rankings, and recommendations in unjust ways.

User-friendly interfaces and decentralized applications (apps) that make it simple to publish, browse, search, review, suggest, curate, verify, and data items, vendors, and customers will be required for a successful marketplace. People and organizations should be sufficiently incentivized to build and supply functional and user-friendly apps while keeping them decentralized.

Nonetheless, the smart economy's most pressing difficulty is privacy and security concerns. The smart economy idea, in particular, presents difficulties in analyzing vulnerabilities and revising strategies. Security concerns are real and must be addressed to guarantee the sustained achievement of smart city goals. Since security is a costly operation requiring a large budget, processing in the public sector takes longer. As a result, security and privacy are essential subjects, particularly in smart economy transactions, which are becoming more vital for believing the smart city notion of enhancing living standards. As a result, secure economic services are required to provide quick assistance for transporting data, mainly via smart city networks. Therefore, it is necessary to have secured financial services to extend fast support for moving data, especially over smart city networks. The current research [3] aims to briefly present the core concepts of security and privacy issues concerning smart cities and reveal contemporary cyber-attacks targeting smart cities based on modern literature. Further, this research has elaborated and identified numerous security weaknesses and privacy challenges about various cyber security issues, challenges, and recommendations to provide future directions.

To address the most persistent difficulty in the smart economy area, this article [121] provides a resilient architecture that defends smart city communications using autonomic computing and Moving Target Defense (MTD) approaches. The basic concept behind attaining robustness is to make it exceedingly difficult for attackers to find out the current active execution environments utilized to operate smart city services by randomizing the utilization of these resources at runtime. The authors analyzed and verified their technique by running a variety of assaults on a smart infrastructure testbed and demonstrating that the delivered services could withstand these attacks with minimum overhead.

Under the same assumption, this research [7] focuses on the relevance of mobility and the strength of elliptic curve cryptography in heterogeneous wireless sensor networks to secure smart city applications. The results of the various simulations revealed that the proposed Dynamic Approach to Secure Smart City Applications algorithm improves application security by lowering the energy consumption, the number of calculations required, and the storage space necessary for the elliptic curve cryptography keys. In particular, in the proposed approach, a powerful mobile cluster head performs operations that demand more computation and use more battery sensors regularly, such as the development, maintenance, and distribution of elliptic curve cryptography keys and periodic rekeying. Because the powerful mobile cluster head has no energy limits and is considered impenetrable, it serves as a periodic distributor of cryptographic keys to all the data sensors in the different applications.

However, the internet-based economy creates several threats, such as unlawful data copying and digital copyright infringement, posing the problem of preserving the security and copyright of important data housed in databases. In this study [122], we present a unique approach, Lossless Database Watermarking in Homomorphic Encryption domain (HOPE-L), to solve this difficulty. Our method combines database encryption (homomorphic encryption, order-preserving encryption) and information concealing (lossless databased watermarking) technologies. To be more explicit, the information concealing technique uses the homomorphic encryption algorithm's homomorphic features to incorporate hidden data. Throughout the data embedding process, no distortion is introduced. As a result, we may combine the watermark within the encrypted data without losing any information. The watermark may validate the copyright, and the receivers can retrieve the original database without losing any data. Theorem and analysis show that HOPE-L can attain additional embedding space without distortion. Extensive testing demonstrates that the operating method is time-efficient, the embedded watermark is resilient, and HOPE-L outperforms previous techniques and can withstand typical database assaults.

In contrast, the creative contribution of Blockchain to smart economy activities may enhance corporate operations and provide a real-time view of all financial data. On-demand products, for example, of a signed contract in the shortest period feasible may considerably assist financial operations. This process's approval time is too lengthy, creating an opportunity to use Blockchain as a solution. This project [123] focuses on the Document of Understanding (DOU) contract, which serves as the foundation for the relationship between a consumer service and the supplier of that service. The authors leveraged local resources and used design thinking and agile principles to construct a local Blockchain Ledger for this assignment. As a consequence of merging the smart economy with cutting-edge technology, company operations in Smart Cities may be strengthened by safely using smart contracts in day-to-day activities. As a result, they produced a proof-of-concept Blockchain demo that has the whole precise history of the agreement, with immutable transactions and transparency, while providing protection and privacy to the participant's information. They registered the time required to get the DOU contract signed off by everyone engaged in the process with this demo, significantly improving it. It may also be reproduced in other areas that handle sensitive data and financial reporting.

It must be noted, that Advanced Persistent Threat (APT) attacks [124] are one of the significant security challenges confronting smart economy sectors in smart cities. APT is a stealthy threat actor that acquires unauthorized network access and stays undiscovered for a lengthy period. The targets of these meticulously selected and studied attacks are often financial institutions or governmental networks. According to this viewpoint, high-level security requirements must be enforced as smart economy frameworks, and infrastructures grow more technology-dependent. Furthermore, the candidates must work hard in the front, keeping an eye out for suspicious activity and aberrant conduct. To guarantee compliance with security rules and regulations, measures are essential to secure the weakest link in the smart economy IT infrastructure - endpoints and end-user devices.

*2.7 Smart People*

Smart People's domain aspires to change how people engage with the public and commercial sectors as individuals or enterprises via information or service supply. Increasing social and digital inclusion/equality via educational offerings is necessary for delivering more efficient communication and services based on new technologies. Also, smart people are about smart types of education that promote job options, labor market possibilities, vocational training, and lifelong learning for people of all ages and ethnicities. Talent development is also crucial from an economic development standpoint since it is becoming an increasingly relevant location element. Smart solutions for smart people promote the creation of a welcoming and inclusive atmosphere to boost prosperity and creativity within a city or community. Implementing intelligent solutions facilitates or fosters participation, open-mindedness, and innovation. In general, a city is smart because it leverages technology to improve the lives of its residents [1].

Only humans can use technology, improve economic and political efficiency, and contribute to social, cultural, and urban advancement. However, poor morale and intellect, a lack of qualified human resources, and multi-ethnic conflicts are vital difficulties that often contribute to societal issues. This study [125] aimed to discover and investigate the aspects of clever people for a smart city. The research used a mixed-method approach. Questionnaires, document reviews, and observations were used to gather data. The findings revealed that the variables of agreeableness, conscientiousness, emotional stability, extraversion, and experience with openness all had high mean scores. Aside from that, the component of friendliness had the highest average mean score of 3.78 out of the four factors: conscientiousness, extraversion, emotional stability, and experience with openness. This research's findings indicate that local governments must adopt strategies and policies to construct and promote smart cities.

Currently, schools disperse the teaching of various components of data skills throughout the curriculum. However, as smart city technologies emerge and demonstrate real promise in contributing to a more sustainable future, it is clear that new skills for working with the large urban data sets that drive these innovations must be taught to future generations for them to be active smart city citizens. The authors of this study [126] question how data skills might be taught more cohesively and practically, allowing for the application of skills in actual, smart city scenarios. They suggest using Urban Data Games to provide a setting for learning and showing practical application of skills for dealing with substantial, complicated data sets, such as big data on smart home energy use. In this regard, are current study programs capable of meeting the need for smart city education? What are the opportunities for people to become smart in a smart city? This study [127] offers excellent integration of smart people educational programs in future smart cities based on practice and professional field-oriented, diversity-inclusive approach. As, it is commonly acknowledged that learning-by-doing may considerably improve students' understanding of information security in smart cities. Hands-on laboratories may help individuals learn about security fundamentals. This study [128] includes various hands-on laboratories that might assist individuals in doing practical exercises in risk-free settings.

But low motivation is the first barrier when educating end-users about smart cities. Game-based learning with interactive exercises and engaging multimedia is an effective way to motivate end-users. Providing a wide range of game material to meet educational demands is critical. For example, in this paper [129], the authors propose a phishing attack game to describe stereotypical features of phishing attack techniques to teach people. As anti-phishing games develop as a scalable, motivating, and practical way to anti-phishing education for non-professional end-users, issues occur owing to a game's content and context becoming irrelevant. When a game delivers unfamiliar or unrelated instances to the user, the learning potential is restricted since the user lacks a reference point. This study [130] presents a customization pipeline for data collecting, creation, and distribution for anti-phishing learning games to give players more meaningful, relevant game material. With the rise of remote work and education, it is vital to adopt new technologies to teach cybersecurity ideas. This work [131] describes the concept, design, and prototype of mixed reality-based cybersecurity teaching application on phishing to expose schoolchildren to the topic remotely and allow them to practice distinguishing harmful from authentic mail. On the other hand, there are methods systems to learn without much input from humans, e.g., phishing websites can be detected using machine learning by classifying the websites as legitimate or illegitimate [132]. Furthermore, this paper [133] describes a contemporary research group's attempt to counteract targeted assaults using spear-phishing by using social engineering via user education (to increase the success probability of phishing attacks, attackers often adopt social engineering techniques). The authors specifically establish a link between human psychological features and sensitivity to social engineering. The outcome may be used to determine if a user has been exposed to a social engineering approach, and the outcome can be used for countermeasure or user training.

On the other side, ICT can potentially increase people's intelligence. This paper [134] describes a system that employs smart plugs, smart cameras, smart power strips, and a digital assistant such as Amazon Alexa, Google Home, Google Assistant, Apple Siri, or Microsoft Cortana to capture voice commands spoken in a much more natural manner by a person with physical disabilities to control ordinary home electrical appliances to turn them on or off with minimal effort. Moreover, this research [135] intends to assist blind persons using smartphone devices. The program allows users to start any app and call contacts using voice commands. Speech commands may be used to instruct a mobile device. These orders are quickly interpreted by the voice recognition engine, which transforms speech into text for direct actions. This strategy is beneficial when a person feels alone in a low setting since it allows him to make a voice call to a known individual. Aside from that, the system offers an app interface that allows the user to get the most recent information from numerous web servers.

To allow smart connections among various devices, smart city technologies have merged artificial intelligence into smart gadgets. AI-powered smart home gadgets may interact with one another and collect new data to aid in learning human routines. The information gathered is utilized to forecast user behavior and establish situational awareness, i.e., to comprehend user preferences and modify settings appropriately. While we recognize some ideals critical for the ethical discussion in AI, such as fairness, transparency, and accountability, these values often clash. More openness, for example, may result in less privacy. Introducing higher principles to balance values raises two issues:

Principles might also be in contradiction with one another, deflecting the issue into a purely speculative sphere.

If a higher-level principle is presented and contradicts another, a higher-higher-level principle is required, and we shall enter an endless regress.

Although AI ethics is part of the so-called field of applied ethics, it appears to be about applying principles and values and finding the right balance concerning specific ethical theories, such as Kantian ethics or utilitarianism, traditional approaches in the field of applied ethics do not provide sufficient conceptual means to deal with practical problems. As a result, the difficulties that occur from installing intelligent systems cannot be fully addressed since the same issue arises: How may values be balanced concerning ethical theories? If higher-level principles are not feasible for resolving value conflicts, the conditions under which they may be applied should be considered. Consequently, it is advocated that precise criteria for implementing the principles be made clear, resulting in at least a clarification for public discussion regarding some technical developments in AI. Furthermore, it is argued that to address these conflicts, the implementation must be reviewed to see whether it allows for future human involvement rather than rendering future actions and interventions impossible [136].

Our degree of connectedness and the rate of technology changes may be perplexing. It is widely accepted we've arrived at a crossroads at which we must reconsider how we engage with one another, what we value, how we work, and how we see ourselves. Recognizing the need for us all to look up from our devices, disconnect, breathe fresh air, enjoy some green space, and engage in face-to-face conversations is part of thoughtful, smart city development. Privacy, consumerism, ownership, and human connections are just a few significant concerns that are touched on. Cities are fundamental to how the world operates. Keeping outstanding talent in our cities will provide possibilities we cannot afford to miss.

## 3. Discussion and Conclusion

The Smart Cities paradigm emerges as a reaction to the objective of constructing the city of the future, where inhabitants' and industry well-being and rights are secured, and urban planning is evaluated from an environmental and sustainable standpoint. The development of smart networks must invariably involve the provision of integrated cyber and privacy ICT solutions [59], [137]. These solutions must ensure the interoperability of the various elements that make up the city's structure and lessen the likelihood that multiple technologies will become obsolete. The variety of the infrastructures and the dynamism of their operational environment necessitates a continual reduction in complexity, quicker processing of expansion works, and the inclusion of equivalent new ones [3], [12]. These requirements must be met. In addition, unified management proposes clear and definite ways of providing end-to-end smart services based on robust security standards and ensuring the privacy of the information being exchanged to offer quality services [139], [140]. These ways are based on the fact that smart cities are becoming increasingly interconnected.

In this particular piece of work, to assess the developing dangers, some specific events of threats, attacks, and their respective countermeasures have been selected given. These are the kind of occurrences that have been suggested from time to time in the scientific literature. The work seeks to be an indicative model that may be considered during the architectural design and execution of infrastructure improvements connected to intelligent city networks. This goal was set for the work.

Future enhancements will include incorporating operational standards used in industrial network applications, subject to ongoing modification and reordering, and newly recognized standards for smart city networks. Also, the recording of the general recommendations by the standardizing bodies per field of operation of the smart networks, as well as the corresponding gaps that have possibly been identified and further concern development and evaluation procedures, is a significant development that bears mentioning. This is another important advancement.

21[13] N. V. Lopes, "Tutorial: Smart Governance for Smart Cities," in *2018 International Conference on eDemocracy & eGovernment (ICEDEG)*, Apr. 2018, pp. 1–2. doi: 10.1109/ICEDEG.2018.8372349.

[14] N. V. Lopes, "Smart governance: A key factor for smart cities implementation," in *2017 IEEE International Conference on Smart Grid and Smart Cities (ICSGSC)*, Jul. 2017, pp. 277–282. doi: 10.1109/ICSGSC.2017.8038591.

[15] A. Roy, "Smart delivery of multifaceted services through connected governance," in *2019 3rd International Conference on Computing Methodologies and Communication (ICCMC)*, Mar. 2019, pp. 476–482. doi: 10.1109/ICCMC.2019.8819851.

[16] A. Ruiz-Zafra, J. Pigueiras, A. Millán-Alcaide, V. M. Larios, and R. Maciel, "A Digital Object-based Infrastructure for Smart Governance of Heterogeneous Internet of Things Systems," in *2020 IEEE International Smart Cities Conference (ISC2)*, Sep. 2020, pp. 1–8. doi: 10.1109/ISC251055.2020.9239077.

[17] Y. Guo, Z. Tang, P. Li, and H. Peng, "The research on the connotation characteristics and policy analysis of smart urban agglomeration under holistic governance," in *2021 International Conference on Big Data and Intelligent Decision Making (BDIDM)*, Jul. 2021, pp. 96–99. doi: 10.1109/BDIDM53834.2021.00026.

[18] S. Bartolucci and S. Fiorentino, "Blockchain and Smart Contracts as New Governance Tools for the Sharing Economy," in *2021 IEEE 18th International Conference on Software Architecture Companion (ICSA-C)*, Mar. 2021, pp. 118–119. doi: 10.1109/ICSA-C52384.2021.00030.

[19] M. Razaghi and M. Finger, "Smart Governance for Smart Cities," *Proc. IEEE*, vol. 106, no. 4, pp. 680–689, Apr. 2018, doi: 10.1109/JPROC.2018.2807784.

[20] J. Gowthami, N. Shanthi, and N. Krishnamoorthy, "Secure Three-Factor Remote user Authentication for E-Governance of Smart Cities," in *2018 International Conference on Current Trends towards Converging Technologies (ICCTCT)*, Mar. 2018, pp. 1–8. doi: 10.1109/ICCTCT.2018.8551172.

[21] Z. Tang, "Analysis of information security problems and countermeasures in big data management of colleges and universities under smart campus environment," in *2021 2nd International Conference on Information Science and Education (ICISE-IE)*, Aug. 2021, pp. 912–915. doi: 10.1109/ICISE-IE53922.2021.00209.

[22] D. E. MAJDOUBI, H. EL BAKKALI, and S. SADKI, "Towards Smart Blockchain-Based System for Privacy and Security in a Smart City environment," in *2020 5th International Conference on Cloud Computing and Artificial Intelligence: Technologies and Applications (CloudTech)*, Aug. 2020, pp. 1–7. doi: 10.1109/CloudTech49835.2020.9365905.

[23] H. Alamleh and A. A. S. AlQahtani, "Analysis of the Design Requirements for Remote Internet-Based E-Voting Systems," in *2021 IEEE World AI IoT Congress (AIIoT)*, Feb. 2021, pp. 0386–0390. doi: 10.1109/AIIoT52608.2021.9454194.

[24] J. Thakkar, N. Patel, C. Patel, and K. Shah, "Privacy-Preserving E-voting System through Blockchain Technology," in *2021 IEEE International Conference on Technology, Research, and Innovation for Betterment of Society (TRIBES)*, Sep. 2021, pp. 1–6. doi: 10.1109/TRIBES52498.2021.9751618.

[25] Y. Khlaponin, V. Vyshniakov, M. Prygara, and V. Poltorak, "The New Concept of Guaranteeing Confidence in the E-Voting System," in *2020 IEEE International Conference on Problems of Infocommunications. Science and Technology (PIC S&T)*, Jul. 2020, pp. 747–752. doi: 10.1109/PICST51311.2020.9468012.

[26] M. Derawi, Y. Dalveren, and F. A. Cheikh, "Internet-of-Things-Based Smart Transportation Systems for Safer Roads," in *2020 IEEE 6th World Forum on Internet of Things (WF-IoT)*, Jun. 2020, pp. 1–4. doi: 10.1109/WF-IoT48130.2020.9221208.

[27] D. Vishal, H. S. Afaque, H. Bhardawaj, and T. K. Ramesh, "IoT-driven road safety system," in *2017 International Conference on Electrical, Electronics, Communication, Computer, and Optimization Techniques (ICEECCOT)*, Sep. 2017, pp. 1–5. doi: 10.1109/ICEECCOT.2017.8284624.

[28] A.-E. M. Taha, "An IoT Architecture for Assessing Road Safety in Smart Cities," *Wirel. Commun. Mob. Comput.*, vol. 2018, p. e8214989, Nov. 2018, doi: 10.1155/2018/8214989.

25[80] S. K. Das, "Cyber-physical-social convergence in smart living: Challenges and opportunities," in *2016 IEEE International Conference on Pervasive Computing and Communication Workshops (PerCom Workshops)*, Mar. 2016, pp. 1–1. doi: 10.1109/PERCOMW.2016.7457093.

[81] A. J. Hussain, D. M. Marcinonyte, F. I. Iqbal, H. Tawfik, T. Baker, and D. Al-Jumeily, "Smart Home Systems Security," in *2018 IEEE 20th International Conference on High Performance Computing and Communications; IEEE 16th International Conference on Smart City; IEEE 4th International Conference on Data Science and Systems (HPCC/SmartCity/DSS)*, Jun. 2018, pp. 1422–1428. doi: 10.1109/HPCC/SmartCity/DSS.2018.00235.

[82] K. Karimi and S. Krit, "Smart home-Smartphone Systems: Threats, Security Requirements and Open research Challenges," in *2019 International Conference of Computer Science and Renewable Energies (ICCSRE)*, Jul. 2019, pp. 1–5. doi: 10.1109/ICCSRE.2019.8807756.

[83] F. James, "A Risk Management Framework and A Generalized Attack Automata for IoT based Smart Home Environment," in *2019 3rd Cyber Security in Networking Conference (CSNet)*, Jul. 2019, pp. 86–90. doi: 10.1109/CSNet47905.2019.9108941.

[84] A. M. Gamundani, A. Phillips, and H. N. Muyingi, "An Overview of Potential Authentication Threats and Attacks on Internet of Things(IoT): A Focus on Smart Home Applications," in *2018 IEEE International Conference on Internet of Things (iThings) and IEEE Green Computing and Communications (GreenCom) and IEEE Cyber, Physical and Social Computing (CPSCom) and IEEE Smart Data (SmartData)*, Jul. 2018, pp. 50–57. doi: 10.1109/Cybermatics_2018.2018.00043.

[85] M. Ibrahim and I. Nabulsi, "Security Analysis of Smart Home Systems Applying Attack Graph," in *2021 Fifth World Conference on Smart Trends in Systems Security and Sustainability (WorldS4)*, Jul. 2021, pp. 230–234. doi: 10.1109/WorldS451998.2021.9514050.

[86] S. ur Rehman and V. Gruhn, "An approach to secure smart homes in cyber-physical systems/Internet-of-Things," in *2018 Fifth International Conference on Software Defined Systems (SDS)*, Apr. 2018, pp. 126–129. doi: 10.1109/SDS.2018.8370433.

[87] W. She, Z.-H. Gu, X.-K. Lyu, Q. Liu, Z. Tian, and W. Liu, "Homomorphic Consortium Blockchain for Smart Home System Sensitive Data Privacy Preserving," *IEEE Access*, vol. 7, pp. 62058–62070, 2019, doi: 10.1109/ACCESS.2019.2916345.

[88] T. Li *et al.*, "Your Home is Insecure: Practical Attacks on Wireless Home Alarm Systems," in *IEEE INFOCOM 2021 - IEEE Conference on Computer Communications*, Feb. 2021, pp. 1–10. doi: 10.1109/INFOCOM42981.2021.9488873.

[89] W. He *et al.*, "SoK: Context Sensing for Access Control in the Adversarial Home IoT," in *2021 IEEE European Symposium on Security and Privacy (EuroS&P)*, Sep. 2021, pp. 37–53. doi: 10.1109/EuroSP51992.2021.00014.

[90] A. Singh and B. Sikdar, "Adversarial Attack for Deep Learning Based IoT Appliance Classification Techniques," in *2021 IEEE 7th World Forum on Internet of Things (WF-IoT)*, Jun. 2021, pp. 657–662. doi: 10.1109/WF-IoT51360.2021.9594946.

[91] K. Demertzis, N. Tziritas, P. Kikiras, S. L. Sanchez, and L. Iliadis, "The Next Generation Cognitive Security Operations Center: Adaptive Analytic Lambda Architecture for Efficient Defense against Adversarial Attacks," *Big Data Cogn. Comput.*, vol. 3, no. 1, Art. no. 1, Mar. 2019, doi: 10.3390/bdcc3010006.

[92] U. T. Khan and M. F. Zia, "Smart city technologies, key components, and its aspects," in *2021 International Conference on Innovative Computing (ICIC)*, Aug. 2021, pp. 1–10. doi: 10.1109/ICIC53490.2021.9692989.

[93] H. K. Sharma, A. Kumar, S. Pant, and M. Ram, "1 Introduction to Smart Healthcare and Telemedicine Systems," in *Artificial Intelligence, Blockchain and IoT for Smart Healthcare*, River Publishers, 2022, pp. 1–12. Accessed: Jul. 10, 2022. [Online]. Available: https://ieeexplore.ieee.org/document/9782809

[94] F. Hussain and M. Hammad, "Smart healthcare facilities via IOT in the healthcare industry," in *3rd Smart Cities Symposium (SCS 2020)*, Sep. 2020, vol. 2020, pp. 386–391. doi: 10.1049/icp.2021.0945.

[95] H. K. Sharma, A. Kumar, S. Pant, and M. Ram, "3 Role of Artificial Intelligence, IoT and Blockchain in Smart Healthcare," in *Artificial Intelligence, Blockchain and IoT for Smart Healthcare*, River Publishers, 2022, pp. 25–36. Accessed: Jul. 10, 2022. [Online]. Available: https://ieeexplore.ieee.org/document/9782803

28